\documentclass[12pt]{article}
\usepackage{graphicx}

\newcommand\pubnumber{SNSN-323-63}
\newcommand\pubdate{\today}

\def\napoli{Department of Physics, Oregon State University, \\Corvallis, Oregon 97331, USA}
\def\support{\footnote{This work was supported by NSF Award PHY-1505472.}}

\def\Title#1{\begin{center} {\Large #1 } \end{center}}
\def\Author#1{\begin{center}{ \sc #1} \end{center}}
\def\Address#1{\begin{center}{ \it #1} \end{center}}

\newcommand\pubblock{\rightline{\begin{tabular}{l} \pubnumber\\
         \pubdate  \end{tabular}}}
\newenvironment{Abstract}{\begin{quotation}  }{\end{quotation}}
\newenvironment{Presented}{\begin{quotation} \begin{center} 
             PRESENTED AT\end{center}\bigskip 
      \begin{center}\begin{large}}{\end{large}\end{center} \end{quotation}}

\def\beq{\begin{equation}}
\def\eeq#1{\label{#1}\end{equation}}
\def\eeqn{\end{equation}}

\def\beqa{\begin{eqnarray}}
\def\eeqa#1{\label{#1}\end{eqnarray}}
\def\eeqan{\end{eqnarray}}

\let\bar=\overbar

\def\Dslash{\not{\hbox{\kern-4pt $D$}}}
\def\dslash{\not{\hbox{\kern-2pt $\del$}}}

\def\msb{{\bar{\ssstyle M \kern -1pt S}}}

\begin{document}
\begin{titlepage}
\pubblock

\vfill
\Title{Neutrino Cross Sections: Status and Prospects }
\vfill
\Author{M. F. Carneiro\support}
\Address{\napoli}
\vfill
\begin{Abstract}
We summarize the current status of accelerator based neutrino cross-section measurements. 
We focus on the experimental challenges while also presenting the motivation for these measurements. 
Selected results are highlighted after a quick description of the current major collaborations working on the field. 
\end{Abstract}
\vfill
\begin{Presented}
NuPhys2017, Prospects in Neutrino Physics\\
Barbican Centre, London, UK,  December 20--22, 2017
\end{Presented}
\vfill
\end{titlepage}
\def\thefootnote{\fnsymbol{footnote}}
\setcounter{footnote}{0}

\section{Introduction}

Neutrino physics is entering a new era of precision measurements and 
cross section measurements play a vital part. We will not discuss in 
details the effect of cross section measurements on systematic uncertainties 
as it has been well described in the literature \cite{Anko,Mosel,Coloma,Huber}, but rather focus on 
the experimental difficulties of these measurements as well as show the 
shortcomings of the current theoretical models describing neutrino-nucleus 
scattering.  We also highlight new results from MINERvA, 
MicroBooNE, NoVA and T2K which have been released or presented in public conferences prior to the 
time of the NuPhys Workshop (December 2017) as well as given a quick 
description of each experiment.

In Section 1, we discuss the common general goals of the program; 
in Section 2 we present the experimental difficulties involved in these 
measurements; Section 3 have a quick description of the effects for 
oscillation experiments; in Section 4 highlight new results and in 
Section 5 we summarize and discuss future directions. 

\section{Motivation}
There's no doubt that the measurement of neutrino interactions and it properties 
is one of the current most important topics in particle physics. Their non-zero 
mass is yet to be well understood and it is  one of the few concrete hints of physics 
beyond the standard model. With this in mind the U.S. community has included the 
physics necessary to understand neutrino masses as one of its high priorities. 
Currently this goal is being pursued via the Deep Underground Neutrino 
Experiment (DUNE), an international, long baseline, beam based, neutrino 
oscillations project. At the same time the community in Japan has identified the 
Tokai to Hyper Kamiokande project (T2HK) as its main project for 
the next decade.

To actually make good use of the current planned future neutrino facilities,  we have 
to invest in gaining better knowledge of neutrino-nucleus scattering. Even a small improvement in the current state of the art $(5-10)\%$ errors could greatly reduce the needed run time for five-sigma coverage of some desired measurements. The presence of near detectors in said future facilities do not fully solve the problem of neutrino-nucleus interaction uncertainties, we need to support both theoretical and experimental aspects of the field.

\section{Challenges}
Before jumping into challenges lets describe, in a simplistic way, the basic setup of a lepton-nucleus experiment. An incident neutrino interacts with a heavy nucleus inside the detector. In a neutral current scattering the produced lepton is a neutrino of the same leptonic flavor as of the incident particle, while a charged current interaction will produce the charged partner of the incoming neutrino. 

As the final state lepton escapes the nucleus it leaves behind a hadronic shower that goes through nuclear matter before detection. These so called final state interactions (FSI) can change the angle, energy and charge state of the originally produced hadrons. Occasionally produced pions will be totally absorbed within the nucleus and not detectable in the final state. Produced neutrons can also completely escape detection. There is also a non negligible probability that the initial interaction occurs with a pair of correlated nucleons and a second nucleon is released in the initial interaction. These denominated ``two-particle-two-hole'' (2p2h) events have been recently proven to be quantitatively important in measuring neutrino scattering parameters.

The neutrino flux itself presents a challenge. In contrast to its charged lepton counterpart, the energy of the incident neutrino is not known a priori. The neutrino energy, as well as the primary generated hadronic system, can only be estimated from what is observed in the detector after the above mentioned final state interactions. The flux can be understood as a function of neutrino energy, but there's still no neutrino energy information in a event-by-event basis.

\section{Neutrino cross-section effects in oscillation experiments}
As explained in the last session, the incident neutrino energy is not well known, but it is the initial neutrino energy spectrum that is needed in the extraction of oscillation parameters. What's actually available for the use in a neutrino oscillation experiment is a nuclear model that combine the nuclear effects information and all the energy dependence of all exclusive cross sections. This nuclear model, as well as the best estimate of the incoming neutrino energy spectrum, is the input to the production of Monte Carlo predictions which can then be compared to data to extract oscillation parameters.  

The following illustrative conceptual outline of a two-detector, long-baseline oscillation analysis, lines up the importance of said nuclear model: Reconstruct topology and energy in the Near Detector; Use a nuclear model to infer the neutrino interaction energy; Use geometry differences (and oscillation hypothesis) to predict Far Detector flux; Use the nuclear model and the estimated flux to reconstruct topology and energy in the Far Detector and finally Compare mc and data and test your hypothesis. 1There's clearly a strong dependence of the neutrino-oscillation parameters on neutrino-interaction physics. The energy and configuration of interactions observed in the detector data are the combination of: the energy-dependent neutrino flux; the energy-dependent neutrino-nucleon cross section; and these significant energy-dependent nuclear effects.

\section{Highlighted results}

\subsection{Coherent pion production}
Important for its ability to mimic an oscillation experiment's electron neutrino signal, coherent processes need more study and proper understanding. In the charged current case we have 
\begin{equation}
\nu_{l} + A \rightarrow l^{-} + m^{+} + A 
\end{equation} 
where $m^{\mp} = \pi^{\mp}, K^{mp}, \rho^{mp}, \dots$. This is a channel that needs to be understood and taken into account given the common misidentification of the produced pions as protons. The neutral current case
\begin{equation}
\nu_{l} + A \rightarrow \nu_{l} + m^{0} + A 
\end{equation} 
with $m^{0} = \gamma,\pi^{0}, \rho^{0}, \dots,$ is more critical. Neutral current production of $\pi^{0}$ or $\gamma$ can mimic final-state electrons, an important background for $\nu_{\mu} \rightarrow \nu_{e}$ oscillations. In addition neutrino electron elastic scattering produces photon events that are mostly indistinguishable from those coming from neutral current coherent processes.

\begin{figure}[htb]
\centering
\includegraphics[height=2.5in]{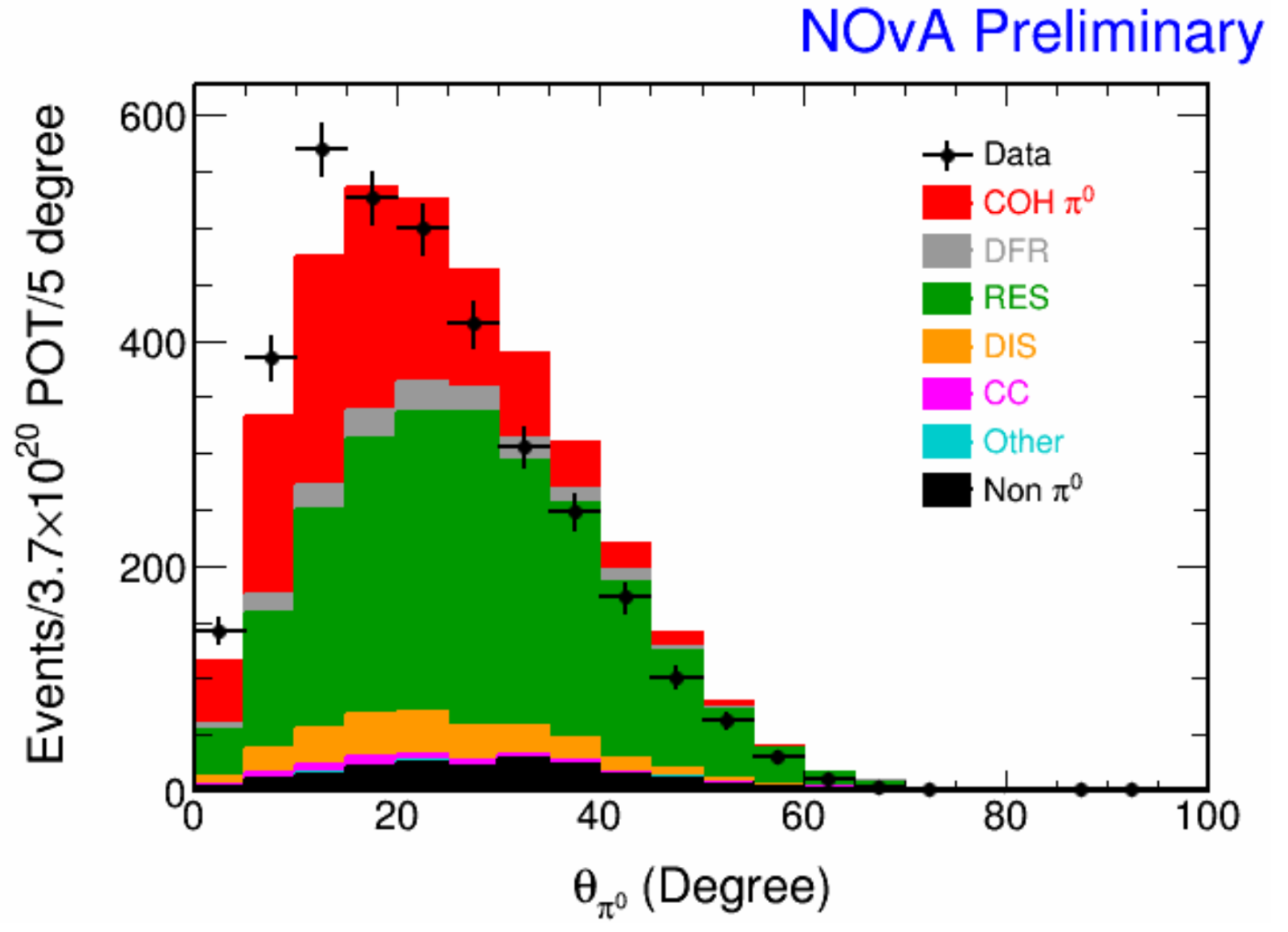}
\caption{NovA preliminary measurement of $\pi^{0}$ angle in respect to the beam direction. Different colors identify Monte Carlo's identification of different processes that contribute to the distribution.\cite{novacoh}}
\label{fig:novacoh}
\end{figure}

\begin{figure}[htb]
\centering
\includegraphics[height=2.5in]{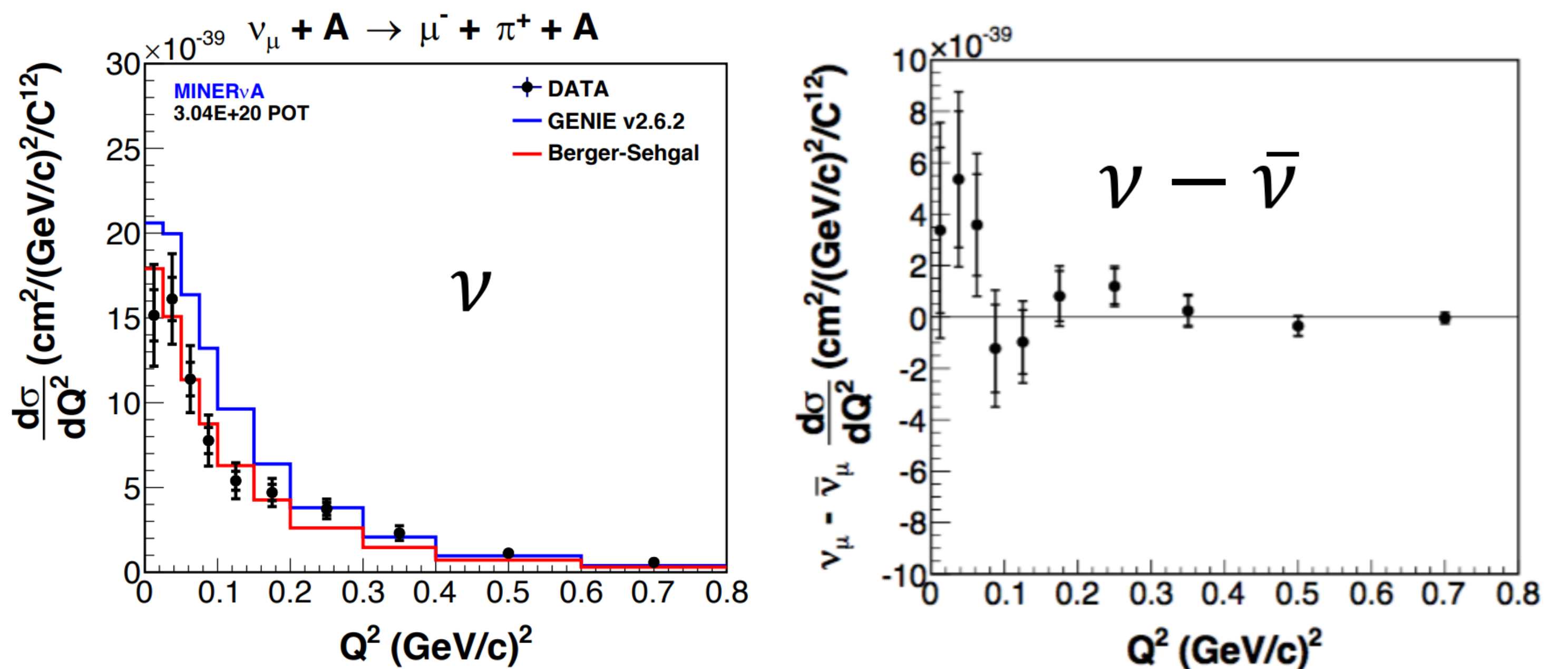}
\caption{Minerva cross section measurement in terms of the transfered energy (left) and difference between neutrino and antineutrino cross section distributions. \cite{coherent minerva}}
\label{fig:minervacoh}
\end{figure}

MINERvA \cite{coherent minerva}, T2K \cite{coherent t2k} and ArgoNeuT \cite{argoneut coherent} have all measured this in charged current interactions. NOvA's near detector design is great for $\pi^{0}$ reconstruction and has searched for this by looking at forward events, preliminary results can be seen in Figure~\ref{fig:novacoh}. This measurements are a powerful check of models that work for charged current \cite{novacoh}. Updated MINERvA results \cite{coherent minerva} include $dE/dQ^2$ and a direct check of the consistency of neutrino and antineutrino cross-section to assess the hypothesis that the process is purely axial and can be seen in Figure~\ref{fig:minervacoh}

\subsection{Charged Current 0 pi}
The CCQE interaction is somewhat better understood but given the final state interactions that the hadronic part undergoes it's impossible to identify true charged current quasi-elastic interactions solely by their topology. Produced protons can undergo several different interactions inside the nuclear matter and also may not have the threshold energy for detection. Pions can be misidentified as protons and neutrons usually escape detection. 
This issue inspired cross section measurement experiments to move to a signal definition anchored in the topology of the final state. The CC0$\pi$ (also referred to as CCQE-like) is defined by a final state that contain the charged lepton produced in the initial interaction, any number of nucleons and no pions.

\subsubsection{Proton muon correlations in CC0$\pi$}
A recent very interesting analysis performed by T2K \cite{t2ktrans} uses events where both the charged lepton and one proton are well reconstructed. In the absence of nuclear effects one would observe conservation of momentum considering the muon and the proton momentum vectors. Transverse variables are defined to study deviations of the transverse momentum from zero to study nuclear effects. $\delta p_{T}$ is defined as the divergence of the transverse momentum conservation and $\delta\alpha_{T}$ as the angle variation of the $\delta p_{T}$ vector.

\begin{figure}[htb]
\centering
\includegraphics[height=2in]{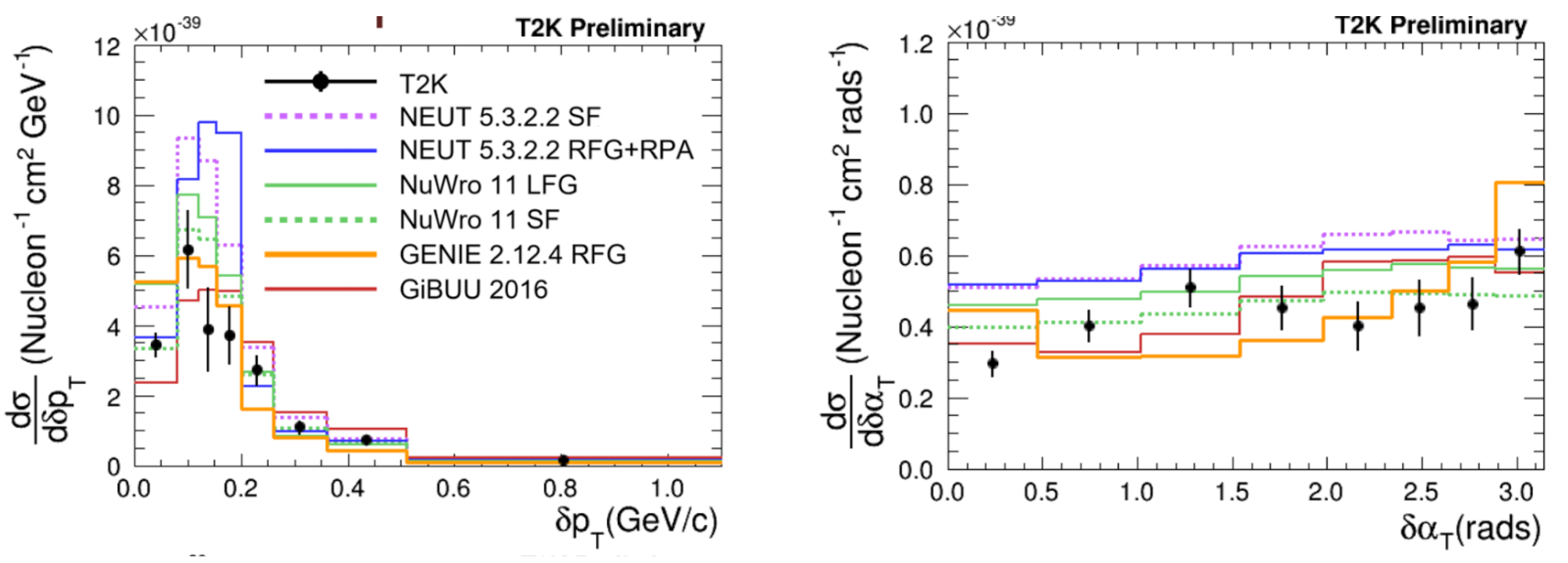}
\caption{Differential cross section measurement in terms of $\delta p_{T}$ defined as the deviation of the transverse momentum from zero (left) and $\delta\alpha_{T}$ as the angle variation of the $\delta p_{T}$ vector, deviation from a flat distribution indicate nuclear effects \cite{t2ktrans}.}
\label{fig:t2ktrans}
\end{figure}

The results (as seen in Figure~\ref{fig:t2ktrans}) compared with default Monte Carlo smulation, done using different event generators, show quite different expectations for the distributions. This analysis can tell us about Fermi motion, 2p2h and help nuclear effect isolation. The study is currently being reproduced in MINERvA \cite{minervatrans}.

\subsubsection{Descriptive CC0$\pi$ model}
Historically the region in the final state hadronic system mass $W$ between the resonance and the quasi-elastic peak is not well modeled. New models added to the standard Monte Carlo, such as RPA (a charge screening nuclear effect) and 2p2h (described in Section 2) improves agreement, but not quite enough. MINERvA uses an inclusive variable $E_{avail}$ to define a 2D Gaussian weight to tune the 2p2h contribution. $E_{avail}$ is defined as the sum of all energy detected apart for neutrons (that scape detection). This tune is designed to empirically fill in the dip region, but not whole kinematic range as it does not scale true QE or resonant production.  

\begin{figure}[htb]
\centering
\includegraphics[height=3.2in]{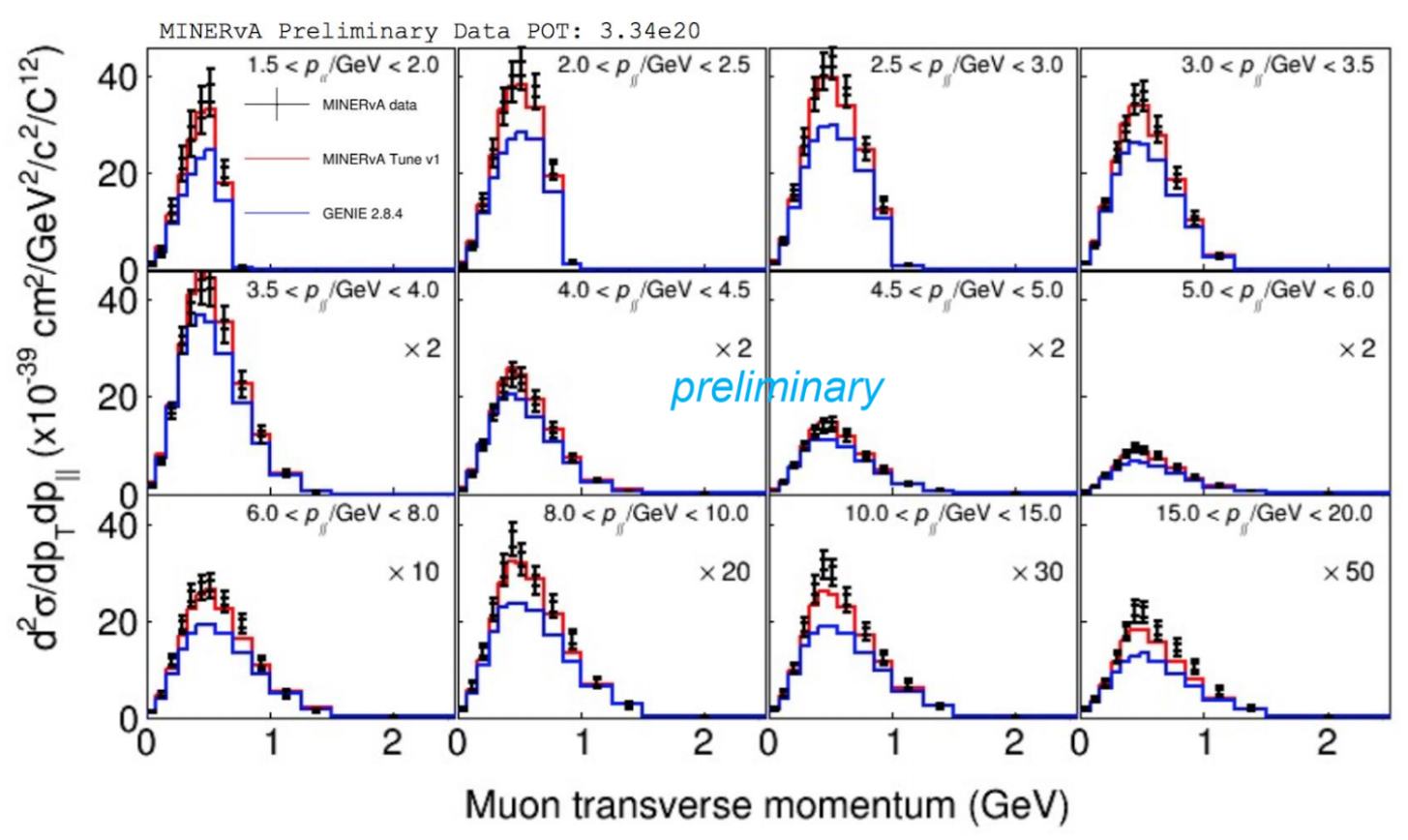}
\caption{Double differential neutrino data in bins of parallel and transversal moments of the leptonic side of the CC0$\pi$ interaction. The blue line show the default Monte Carlo while the red line show the weighted simulation \cite{minervanufit}.}
\label{fig:minervanufit}
\end{figure}

\begin{figure}[htb]
\centering
\includegraphics[height=3.2in]{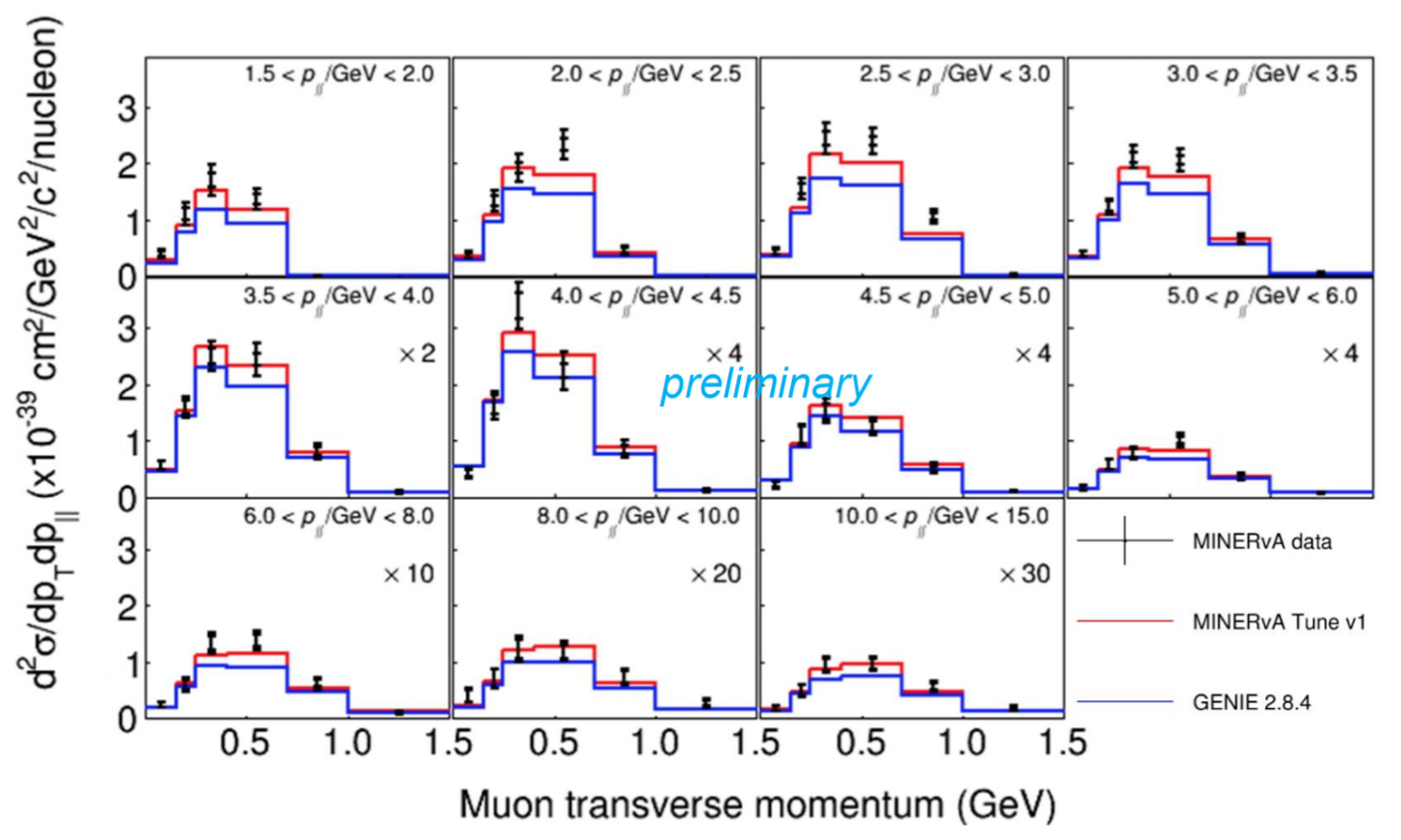}
\caption{Double differential antineutrino data in bins of parallel and transversal moments of the leptonic side of the CC0$\pi$ interaction. The blue line show the default Monte Carlo while the red line show the weighted simulation \cite{minervaanufit}.}
\label{fig:minervaanufit}
\end{figure}

Applying the inclusive fit into the exclusive Double differential neutrino Cross Section improves the data-Monte Carlo agreement.  Figure~\ref{fig:minervanufit} show the distribution for the neutrino case. Impressively, as can be seen in Figure~\ref{fig:minervaanufit}, this inclusive fit done in neutrino data also fits very well the antineutrino data. It worth to note that this fit is not theoretically motivated, but identifies particular energy-momentum transfer and when used to predict other distributions see excellent agreement. It ould be interesting to test this technique against different experimental situations. 

\subsection{Delta resonance in nuclei}
Going up in the energy transfer spectrum from the Quasi-elastic peak we reach the resonance region corresponding to larger hadronic invariant mass. The $\Delta(1232)$ resonance provides  the most important contribution:
\begin{equation}
\nu_{\mu}p \rightarrow \mu^{-}\Delta^{++} ~,~ \Delta^{++}\rightarrow p\pi^{+}.
\end{equation}

\begin{figure}[htb]
\centering
\includegraphics[height=2in]{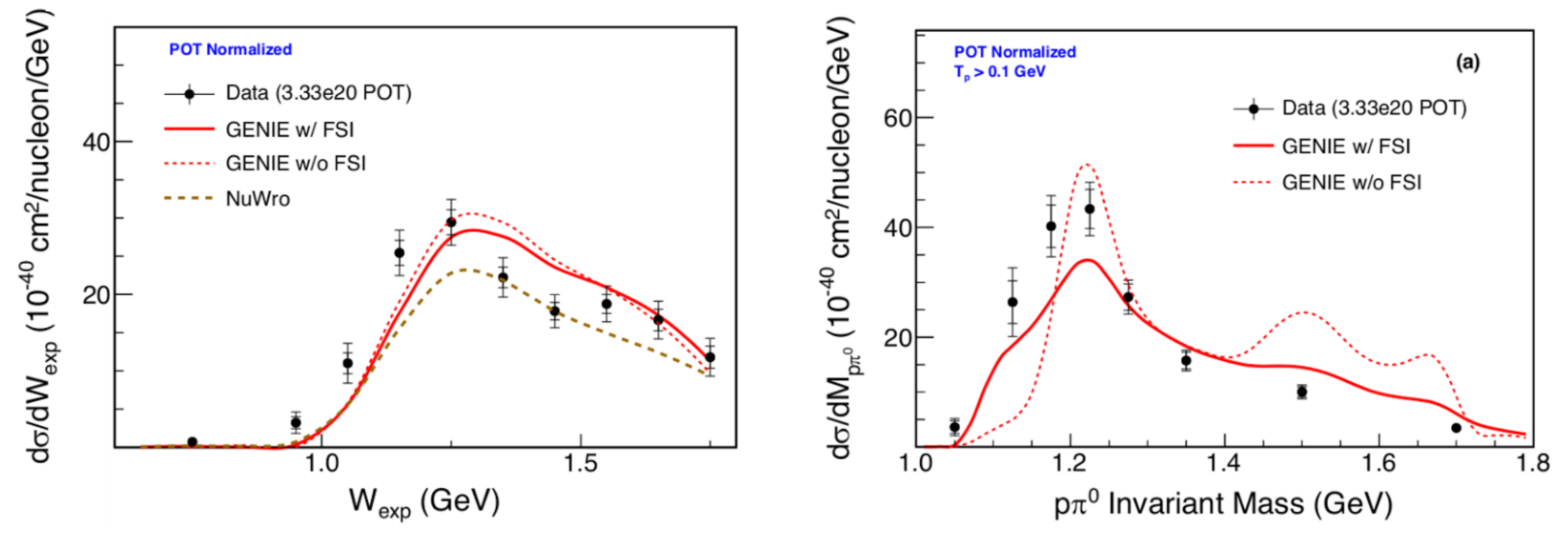}
\caption{Minerva cross-section measurement in terms of measured W (left) and $\pi^{0}$ invariant mass. No current theoretical model implemented in neutrino event generators describe well the data. \cite{minervadelta}}
\label{fig:minervadelta}
\end{figure}

This channel is a important background to the QE process, used as signal by most neutrino oscillation experiments. The main challenge is to isolate pions from proton signals in the detector. Historical tension exists between MINERvA and MiniBooNE data measured on carbon targets (actually CH for MINERvA) \cite{minervaminiboone}. Recent MINERvA results for proton and $\pi^{0}$ final states show some evidence for need of more modern pion models as can be seen in Figure~\ref{fig:minervadelta}.

\subsection{Low threshold multiplicities in LAr}

\begin{figure}[htb]
\centering
\includegraphics[height=2.5in]{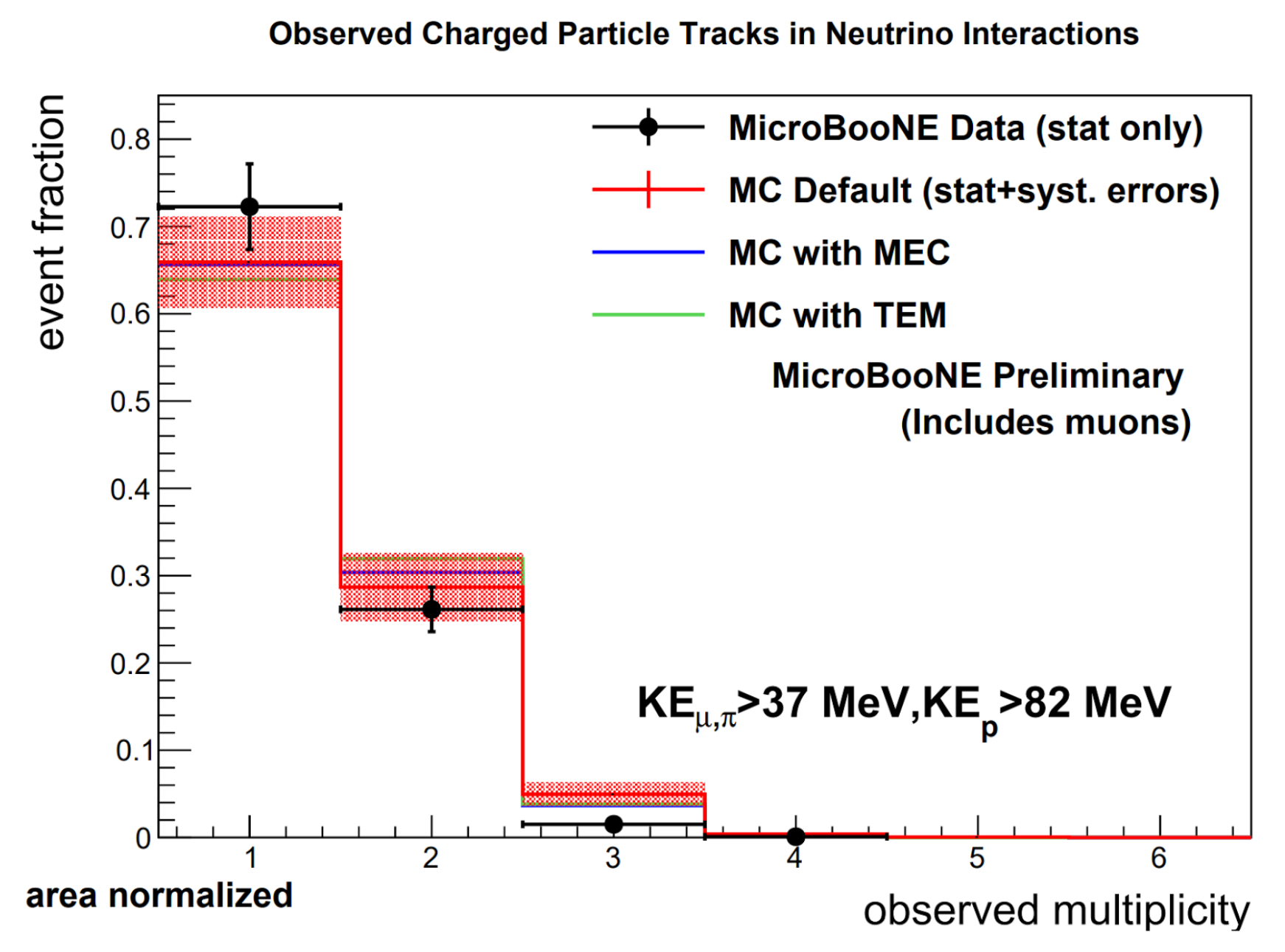}
\caption{Particle multiplicity observed in the MicroBooNE detector \cite{microbmult}}
\label{fig:microbmult}
\end{figure}

MicroBooNE is a very important player in the current scenario as we need more knowledge about interactions in liquid argon. The preliminary result we are highlighting here is an study of the charged multiplicity observed in the detector (Figure~\ref{fig:microbmult}). It works well as a model check of low energy particles, such as spectator nucleons and pions degraded by final state interaction, as well as a good prospect for LAr reconstruction.

\section{Summary}
Neutrino cross section measurements are crucial for oscillation experiments. They are also fascinating physics in their own right. We presented the current experimental and theoretical challenges to these measurements. We need cross section measurements to achieve the expected precision in neutrino oscillation parameters and we will not fully exploit the current planned facilities if there is  not support for the neutrino cross section community. There is a rich field of 
experiments, working with theorists and generator developers ready to meet the challenges.


\end{document}